\documentclass[conference]{IEEEtran}
\IEEEoverridecommandlockouts
\usepackage{bibnames}
\usepackage{amssymb}
\usepackage{verbatim}
\usepackage{amsmath}
\usepackage{amsthm}
\usepackage[pdftex]{graphicx}
\usepackage[font=scriptsize]{caption}
\usepackage{subcaption}
 \usepackage[linesnumbered, ruled,vlined]{algorithm2e}
\usepackage{listings}
\usepackage[flushleft]{threeparttable}
\usepackage{multirow}
\usepackage{listings}
\usepackage{cite}
\usepackage{hyperref}
\usepackage{mdframed}
\usepackage{tabularx}
\usepackage{booktabs}
\usepackage{flushend}
\usepackage{mathrsfs}
\usepackage[T1]{fontenc}
\usepackage[switch]{lineno}
\usepackage{enumitem}

\usepackage{balance}

\definecolor{light-gray}{gray}{0.92}  
  {\begin{mdframed}[backgroundcolor=light-gray]\begin{mdtheorem}{name}{label}}%
  {\end{mdtheorem}\end{mdframed}}

\usepackage[table]{xcolor}
\definecolor{ao}{rgb}{0.0, 0.5, 0.0}
\newtheorem{definition}{Definition}

\newcommand{\etal}{\textit{et al.}\space}

\newcommand{\tool}{\textsc{VFFinder}\xspace}

\newcommand{\aAST}{\textsc{$\alpha$\textit{AST}}\xspace}

\newcommand{\aASTs}{\textsc{$\alpha$\textit{AST}s}\xspace}

\newcommand{\CEL}{CostEffort$@L$\xspace}

\def\BibTeX{{\rm B\kern-.05em{\sc i\kern-.025em b}\kern-.08em
    T\kern-.1667em\lower.7ex\hbox{E}\kern-.125emX}}
\begin{document}

\title{
Silent Vulnerability-fixing Commit Identification Based on Graph Neural Networks
}

\author{\IEEEauthorblockN{Hieu Dinh Vo$^*$\thanks{*Corresponding author.}, Thanh Trong Vu, and Son Nguyen}
\IEEEauthorblockA{\textit{Faculty of Information Technology} \\
\textit{University of Engineering and Technology, Vietnam National University, Hanoi, Vietnam} \\ \{hieuvd,19020626,sonnguyen\}@vnu.edu.vn} 
}
\date{}
\maketitle

\begin{abstract}
The growing dependence of software projects on external libraries has generated apprehensions regarding the security of these libraries because of concealed vulnerabilities. Handling these vulnerabilities presents difficulties due to the temporal delay between remediation and public exposure. Furthermore, a substantial fraction of open-source projects covertly address vulnerabilities without any formal notification, influencing vulnerability management. Established solutions like OWASP predominantly hinge on public announcements, limiting their efficacy in uncovering undisclosed vulnerabilities. To address this challenge, the automated identification of vulnerability-fixing commits has come to the forefront. In this paper, we present \tool, a novel graph-based approach for automated silent vulnerability fix identification. 
%
%
To precisely capture the meaning of code changes, the changed code is represented in connection with the related unchanged code. In \tool, the structure of the changed code and related unchanged code are captured and the structural changes are represented in annotated Abstract Syntax Trees (\aAST). 
\tool distinguishes vulnerability-fixing commits from non-fixing ones using attention-based graph neural network models to extract structural features expressed in \aASTs.
We conducted experiments to evaluate \tool on a dataset of 11K+ vulnerability fixing commits in 507 real-world C/C++ projects. Our results show that \tool significantly improves the state-of-the-art methods by 272--420\% in Precision, 22--70\% in Recall, and 3.2X--8.2X in F1. Especially, \tool speeds up the silent fix identification process by up to 121\% with the same effort reviewing 50K LOC compared to the existing approaches.

\end{abstract}


\section{Introduction}
With the escalating dependence of software projects on external libraries, ensuring their security has emerged as an imperative priority. Vulnerabilities hidden within these libraries can have far-reaching consequences, as exemplified by the infamous Log4Shell\footnote{\url{https://nvd.nist.gov/vuln/detail/CVE-2021-44228}} exploit. 
One critical challenge in addressing these vulnerabilities is the time gap between their fixes and public disclosures~\cite{iso-standard,cert-guideline}. 
For instance, the Log4Shell vulnerability's resolution was introduced four days prior to its public revelation.  
Another illustration involves the Apache Struts Remote Code Execution vulnerability\footnote{\url{https://nvd.nist.gov/vuln/detail/CVE-2018-11776}}, which led to the Equifax breach in 2017, was disclosed in August 2018, but was patched in June 2018\footnote{\url{https://github.com/apache/struts/commit/6e87474}}.
This two-month window provides ample opportunity for the potential exploitation of vulnerable software.
If the library were monitored to identify vulnerability patches, the library's users would have been aware of the potential exploitation and prevented it by updating to the latest version of the component. 

Despite the importance of the vulnerability fix identification task in open-source libraries, only a very small portion of maintainers file for a Common Vulnerability Enumeration (CVE) ID after releasing a fix, while 25\% of open-source projects silently fix vulnerabilities without disclosing them to any official repository~\cite{sspcatcher,snyk-report}. 
This situation raises concerns about the visibility and proactive management of vulnerabilities within the software ecosystem.
The open-source libraries' users rely on several tools and public vulnerability datasets like the Open Web Application Security Project or National Vulnerability Database (NVD). However, CVE/NVD and public databases miss many vulnerabilities~\cite{snyk-report}.
%

%
%

To address this problem, several vulnerability fix identification techniques have been proposed. Following the good practice of coordinated vulnerability disclosure~\cite{iso-standard,cert-guideline}, the related resources of commits, such as commit messages or issue reports, should not mention any security-related information before the public disclosure of the vulnerability. Thus, silent fix identification methods must not leverage these resources to classify commits~\cite{VulFixMiner,Midas,CoLeFunDa}. The state-of-the-art techniques, such as VulFixMiner~\cite{VulFixMiner}, CoLeFunDa~\cite{CoLeFunDa}, and Midas~\cite{Midas}, represent changes in the lexical form of code and apply CodeBERT~\cite{codebert} to capture code changes semantics and determine if they are vulnerability-fixing commit or not. Meanwhile, the existing studies have shown that the semantics of code changes could be captured better in the tree form of code~\cite{fira}.  

This paper, which is the extended version of our previous conference paper \cite{kse}, proposes \tool, a novel graph-based approach for automated vulnerability fix identification. Our idea is to capture the semantic meaning of code changes better, we represent changed code in connection with the related unchanged code and explicitly represent the changes in code structure. 
Particularly, for a commit $c$, the code version before (after) $c$ is analyzed to identify the code necessary for representing the code change, which is the deleted (added) lines of code and their related unchanged lines of code.
After that, the structure of the necessary code of the changes in $c$ is represented by the Abstract Syntax Trees (ASTs). These ASTs are mapped to build an annotated AST (\aAST), a fine-grained graph representing the changes in the code structure caused by $c$. In \aASTs, all AST nodes and edges are annotated \textit{added}, \textit{deleted}, and \textit{unchanged} to explicitly express the changes in the code structure.
To learn the meanings of code changes expressed in \aASTs, we develop a graph attention network model~\cite{gat} to extract semantic features. Then, these features are used to distinguish vulnerability-fixing commits from non-fixing ones. Compared with the previous paper \cite{kse}, this paper provides a more comprehensive example for \aASTs and more experiment results which are conducted on a much larger dataset. 
%


We conducted several experiments to evaluate \tool's performance on a dataset containing 100K+ commits (with 11K+ vulnerability fixing commits) in 507 real-world C/C++ projects. Our results show that \tool significantly improves the state-of-the-art techniques~\cite{Midas,VulFixMiner,jit-fine,jitline} by 272--420\% in Precision, 22--70\% in Recall, and 3.2X--8.2X in F1. Especially, \tool speeds up the silent fix identification process up to 121\%  with the same review effort reviewing 50K lines of code (LOCs) (0.02\% of the total changed LOCs) compared to the existing approaches~\cite{Midas,VulFixMiner,jit-fine,jitline}.

In brief, this paper makes the following major contributions:

\begin{enumerate}
    \item {\tool}: A novel graph-based approach for identifying silent vulnerability fixes.
    \item An extensive experimental evaluation showing the performance of {\tool} over the state-of-the-art methods for vulnerability-fix identification.
\end{enumerate}

The rest of this paper is organized as follows. Section~\ref{sec:representation} describes our novel code change representation. The graph-based vulnerability fix identification model is introduced in Section~\ref{sec:approach}. 
After that, Section~\ref{sec:eval} states our evaluation methodology. Section~\ref{sec:results} presents the experimental results following the introduced methodology and some threats to validity.
Section~\ref{sec:related} provides the related work. Finally, Section~\ref{sec:conclusion} concludes this paper.
\section{Code Change Representation}
\label{sec:representation}

\begin{figure}
    \centering
    \includegraphics[width=0.7\columnwidth]{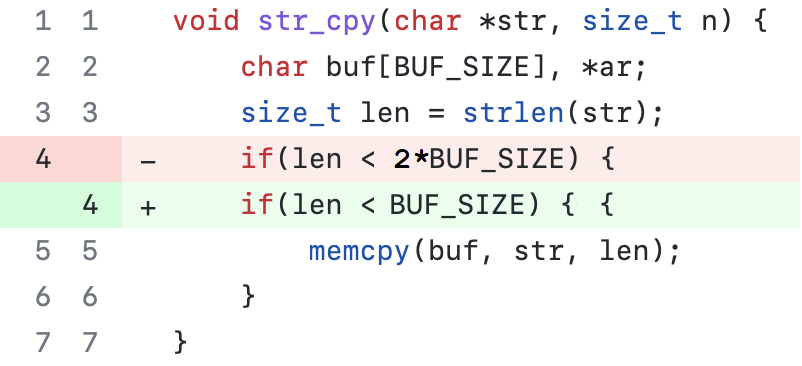}
    \caption{A commit deleting a line of code and adding another to fix a buffer overflow vulnerability}
    \label{fig:example_change}
\end{figure}

Essentially, vulnerability-fixing commits tend to correct the vulnerable code that already exists in repositories. Thus, \textit{to precisely capture the semantics of code changes, besides the changed parts, the unchanged code is also necessary}~\cite{cpathminer,codejit}. Indeed, the unchanged code could connect the related changed parts and help to understand the code changes as a whole.
Additionally, once changed statements are introduced to a program, they change the program's behaviors by interacting with the unchanged statements via certain code relations such as control/data dependencies. Hence, precisely distinguishing similar changed parts could require their relations with the unchanged parts.
In Figure~\ref{fig:example_change}, the unchanged parts, including lines 3 and 5 in both the versions before and after the commit, are necessary for understanding the meaning of the changed code.

\begin{figure*}
    \centering
    \includegraphics[width=1.9\columnwidth]{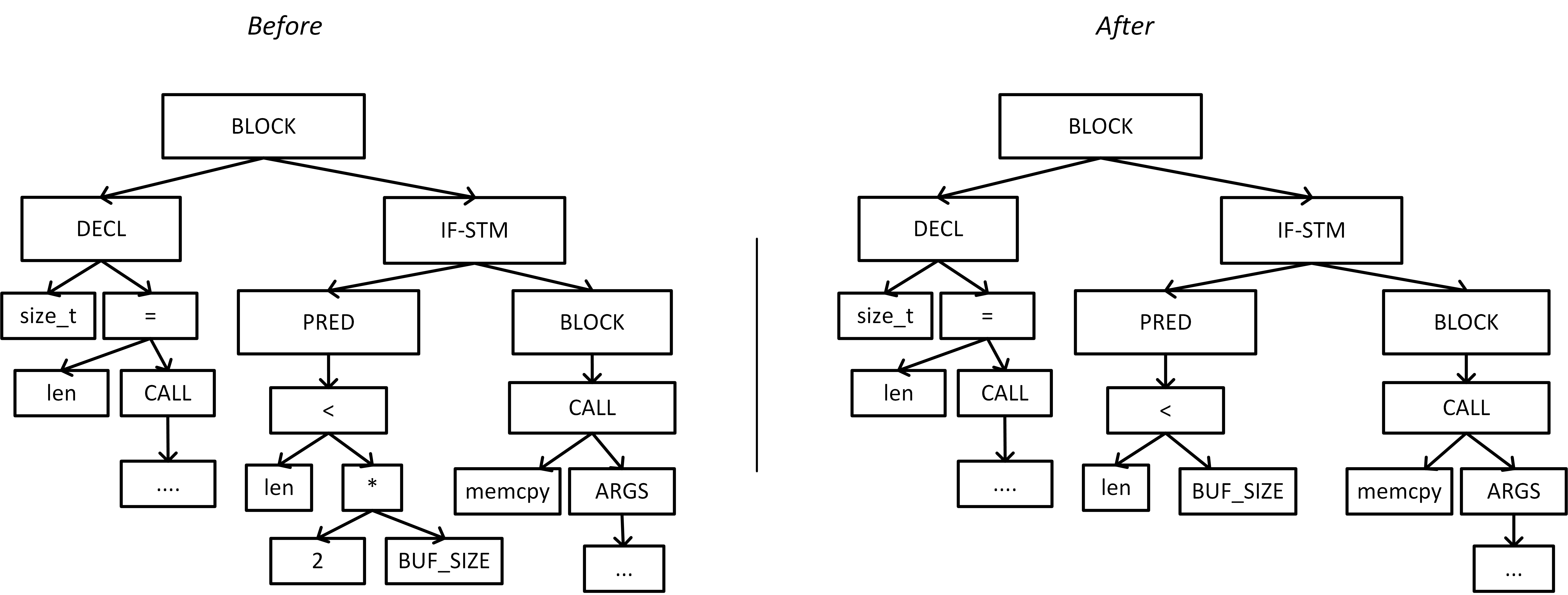}
    \caption{The ASTs of the versions before and after the commit shown in Figure~\ref{fig:example_change}}
    \label{fig:asts}
\end{figure*}  

Thus, for a commit, we additionally consider the code statements that are semantically related to the changed statements in the versions ($v_b$ and $v_a$) before and after the commit. Particularly, the considering statements in $v_b$ are the deleted statements and the related ones via control and data dependencies in $v_b$, while the added statements and their related ones via control and data dependencies in $v_a$ are considered. These statements in $v_b$ and $v_a$ are analyzed to construct corresponding Abstract Syntax Trees (ASTs) and capture the structural changes in an \textit{Annotated AST} - \aAST.
%
%
%
\begin{figure}
    \centering
    \includegraphics[width=1.0\columnwidth]{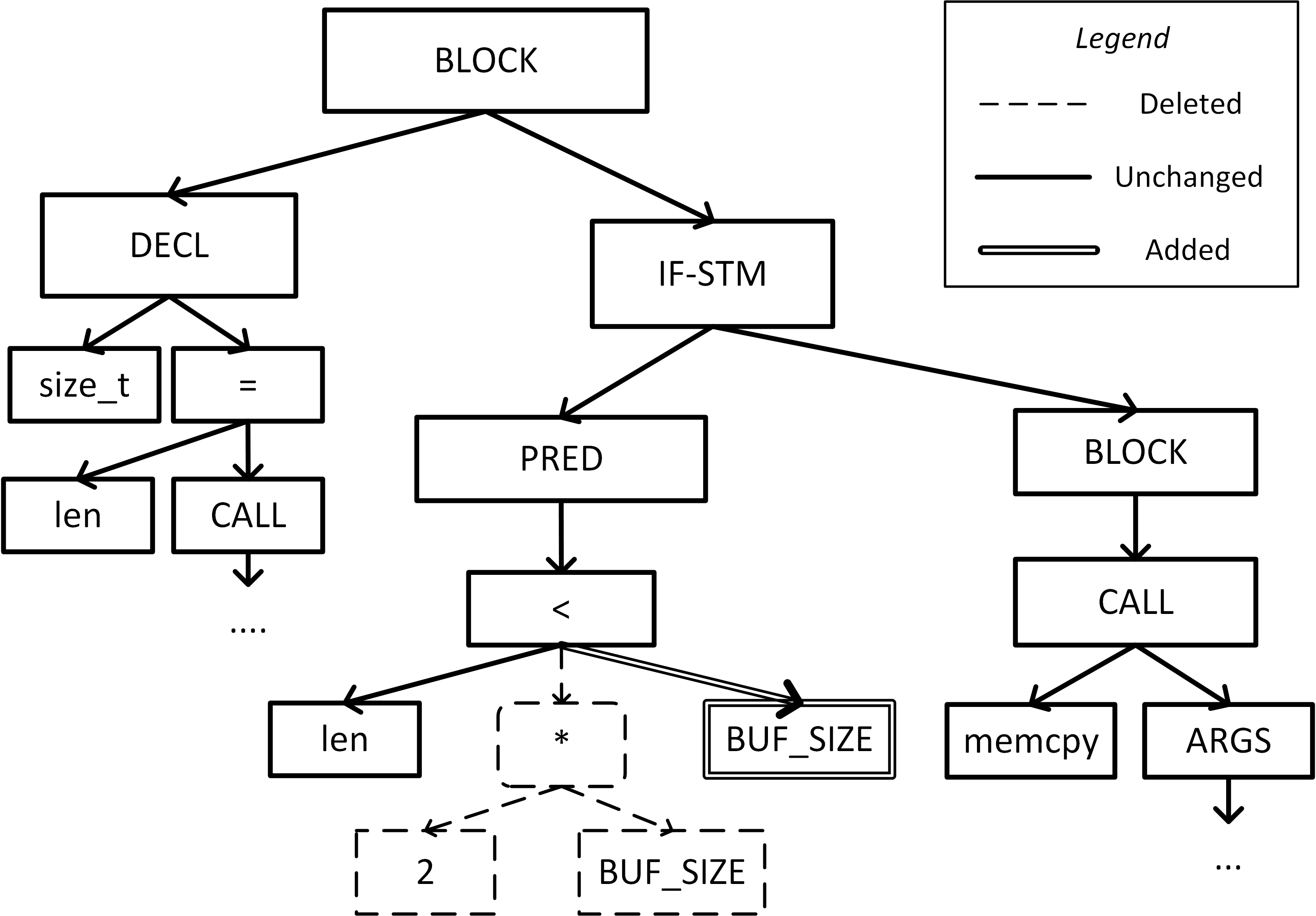}
    \caption{The \aAST corresponding to the commit shown in Figure~\ref{fig:example_change}}
    \label{fig:aAST}
\end{figure}

\begin{definition}[Annotated AST - \aAST]
For a commit changing code from one version ($v_b$) to another ($v_a$), the \textit{annotated abstract syntax tree (annotated AST)} is an annotated graph constructed from the ASTs of $v_b$ and $v_a$. Formally, for $AST_o = \langle N_b, E_b \rangle$ and $AST_{n} = \langle N_a, E_a\rangle$ which are the ASTs of $v_b$ and $v_a$, respectively, the \aAST $\mathcal{T} = \langle \mathcal{N}, \mathcal{E}, \alpha \rangle$ is defined as followings:
\begin{itemize}
    \item $\mathcal{N}$ consists of the AST nodes in the old version and the new version, $\mathcal{N} = N_b \cup N_a$.

    \item $\mathcal{E}$ is the set of the edges representing the structural relations between AST nodes in $AST_o$ and $AST_n$, $\mathcal{E} = E_b \cup E_a$.

    \item Annotations for nodes and edges in $\mathcal{T}$ are either \textit{unchanged}, \textit{added}, or \textit{deleted} by the change. Formally, $\alpha(g) \in \{$\textit{unchanged}, \textit{added}, \textit{deleted}$\}$, where $g$ is a node in $\mathcal{N}$ or an edge in $\mathcal{E}$:
        \begin{itemize}
            \item $\alpha(g) =$ \textit{added} if $g$ is a node and $g \in N_a \setminus N_b$, or $g$ is an edge and $g \in E_a \setminus E_b$
            \item $\alpha(g) =$ \textit{deleted} if $g$ is a node and $g \in N_b \setminus N_a$, or $g$ is an edge and $g \in E_b \setminus E_a$
            \item Otherwise, $\alpha(g) = $ \textit{unchanged} 
        \end{itemize}
\end{itemize}
\end{definition}

Figure~\ref{fig:asts} shows ASTs of the versions before and after the commit shown in Figure~\ref{fig:example_change}. The annotated AST constructed from these ASTs is shown in Figure~\ref{fig:aAST}. The \aAST expresses the change in the structure of the code. Particularly, the right-hand-side of the \texttt{less-than} expression (\texttt{2*BUF\_SIZE}) is replaced by   expression \texttt{BUF\_SIZE}. The \aAST also expresses the structure of the related unchanged statements, which clarify the meaning of the changed code.



\section{Vulnerability-Fix Identification Model}
\label{sec:approach}

\begin{figure*}
    \centering
    \includegraphics[width=2\columnwidth]{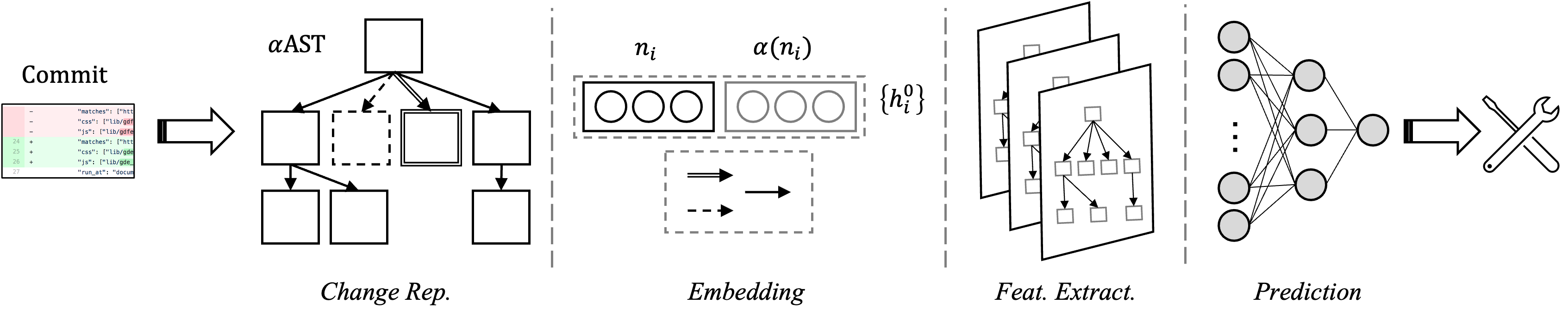}
    \caption{Graph-based Vulnerability-fix Identification Model}
    \label{fig:approach}
\end{figure*}

Figure.~\ref{fig:approach} illustrates the overview of \tool for vulnerability fix identification. First, the given commits and their repositories are used to construct their corresponding \aASTs (\textit{Change representation}). Each AST node in \aASTs is embedded in the corresponding vectors (\textit{Embedding}). After that, a Graph Neural Network (GNN) is applied to extract structural features from constructed \aASTs (\textit{Feature extraction}). Finally, the extracted structural features are used for learning and predicting vulnerability-fixing commits (\textit{Prediction}).

Particularly, in the \textit{Embedding} step, for each \aAST, $\mathcal{T} = \langle \mathcal{N}, \mathcal{E}, \alpha \rangle$, every node in $\mathcal{N}$ is embedded into $d$-dimensional hidden features $n_i$ produced by embedding the content of the nodes. 
Constructing the vectors for nodes' content could be done by applying code embedding techniques~\cite{code2vec, codebert,codet5,word2vec_1, embedding_emse22}.
In this work, we use Word2vec~\cite{word2vec_1}, one of the most popular code embedding techniques for code~\cite{embedding_emse22}. The reason is that the number of AST nodes in \aASTs could be huge. Thus, for a practical embedding step for \aASTs, we apply Word2vec, which is known as an efficient embedding technique~\cite{embedding_emse22}.
Then, to form the node feature vectors, the node embedding vectors are annotated with the change operators (\textit{added}, \textit{deleted}, and \textit{unchanged}) by concatenating corresponding one-hot vector of the operators to the embedded vectors, $h^{0}_i = [n_i || \alpha(n_i)]$, where $||$ is the concatenation operation and $\alpha$ returns the one-hot vector corresponding the annotation of node $i$.

In the \textit{Feature Extraction} step, from each \aAST, $\mathcal{T} = \langle \mathcal{N}, \mathcal{E}, \alpha \rangle$, we develop a Graph Attention Network (GAT)~\cite{gat} model to extract the structural features $H$.
Particularly, the embedded vectors of the nodes from the \textit{Embedding} step are fed to a GAT model.
Each GAT layer computes the representations for the graph's nodes through message passing~\cite{gcn,gat}, where each node gathers features from its neighbors to represent the local graph structure. Stacking $L$ layers allows the network to build node representations from each node's $L$-hop neighborhood.
From the feature vector $h_{i}$ of node $i$ at the current layer, the feature vector $h'_{i}$ at the next layer is: 
$$
h'_{i} =\sigma 
        \left( 
            \sum_{j \in \mathcal{N}_i} \alpha_{ij} \text{\textbf{W}}h_{j}
        \right)
$$
where \textbf{W} is a learnable weight matrix for feature transformation, $\mathcal{N}_i$ is the set of neighbor indices of node $i$ including node $i$ itself via \textit{self-connection}, which is a single special relation from node $i$ to itself.
$\sigma$ is a non-linear activation function such as ReLU. Meanwhile, $\alpha_{ij}$ specifies the weighting factor (importance) of node $j$'s features to node $i$. $\alpha_{ij}$ could be explicitly defined based on the structural properties of the graph or learnable weight~\cite{gcn,fastgcn}. In this work, we implicitly define $\alpha_{ij}$ based on node features~\cite{gat} by employing the self-attention mechanism, where unnormalized coefficients $E_{ij}$ for pairs of nodes $i,j$ are computed based on their features:
$$
E_{ij} = \text{LeakyReLu}(\text{\textbf{a}}^T  \cdot [\text{\textbf{W}}h_i || \text{\textbf{W}}h_j] ), 
$$
where $||$ is the concatenation operation and \textbf{a} is a parametrizing weight vector implemented by a single-layer feed-forward neural network. $E_{ij}$ indicates the importance of node $j$'s features to node $i$. The coefficients are normalized across all choices of $j$ using the softmax function:
$$
\alpha_{ij} = \text{softmax}_j(E_{ij}) = 
\frac{\exp(E_{ij})}{\sum_{k \in \mathcal{N}_i} \exp(E_{ik}) }
$$
After $L$ GAT layers, a $d$-dimensional graph-level vector representation $H$ for the whole CTG $\mathcal{T} = \langle \mathcal{N}, \mathcal{E}, \alpha \rangle$ is built by averaging over all node features in the final GAT layer, $H = \frac{1}{|\mathcal{N}|} \sum_{i \in [1, |\mathcal{N}|]} h^L_i$.
Finally, in the \textit{Prediction} step, the graph features are then passed to a Multilayer perceptron (MLP) to classify if \aAST $\mathcal{T}$ is a fixing commit or not.

\section{Evaluation Methodology}
\label{sec:eval}

To evaluate our vulnerability-fixing commit identification approach, we seek to answer the following research questions:

\noindent\textbf{RQ1: \textit{Accuracy and Comparison}.} How accurate is {\tool} in identifying vulnerability-fixing commits? And how is it compared to the state-of-the-art approaches~\cite{Midas,VulFixMiner}?

\noindent\textbf{RQ2: \textit{Intrinsic Analysis}.} How do the consideration of related unchanged code and the GNN model in {\tool} impact \tool's performance?

\noindent\textbf{RQ3: \textit{Sensitivity Analysis}.} How do various factors of the input, including training data size and changed code complexity, affect {\tool}'s performance?

\noindent\textbf{RQ4: \textit{Time Complexity}.} What is  {\tool}'s running time?

\subsection{Dataset}
In this work, we collect the vulnerability-fixing commits from various public vulnerability datasets~\cite{fixing_database1,bigvul,devign}.
In total, we collected the commits in real-world 507 C/C++ projects, including about 11K fixing commits for the vulnerabilities reported from 1990 to 2022.
%
%
Table~\ref{tab:dataset} shows the overview of our dataset. The details of our dataset can be found at: \textit{\url{https://github.com/UETISE/VFFinder}}.

\begin{table*}[]
\centering
\caption{Dataset statistics}
\label{tab:dataset}
\begin{tabular}{@{}lrrrr@{}}
\toprule
         & \#\textit{Time}   & \#\textit{Fixes}     & \#\textit{Non-fixes}       & \#\textit{Changed LOCs}  \\ \midrule
Training & Aug 1998 -- Mar 2017          & 8,471       & 8,471            & 3,869,454         \\
Testing  & Apr 2017 -- Aug 2022          & 2,828      & 86,395             & 271,182,635         \\ \midrule
Total    & Aug 1998 -- Aug 2022          & 11,299      & 94,866           & 275,052,089         \\ \bottomrule
\end{tabular}
\end{table*}

\subsection{Procedure}
For \textbf{RQ1. Accuracy and Comparison}, we compared \tool against the state-of-the-art vulnerability-fix identification approaches: 

1) \textbf{MiDas}~\cite{Midas} establishes distinct neural networks for varying levels of code change granularity, encompassing commit-level, file-level, hunk-level, and line-level alterations, aligning with their inherent categorization. Subsequently, it employs an ensemble model that amalgamates all foundational models to produce the ultimate prediction.

2) \textbf{VulFixMiner}~\cite{VulFixMiner} and CoLeFunDa~\cite{CoLeFunDa} use CodeBERT to automatically represent code changes and extract features for identifying vulnerability fixes. However, as the implementation of CoLeFunDa has not been available, we cannot compare \tool with CoLeFunDa. This is also the reason that Zhou~\etal was not able to compare MiDas with CoLeFunDa in their study~\cite{Midas}.  

Additionally, we applied the same procedure, adapting the state-of-the-art just-in-time defect detection techniques for vulnerability-fix identification as in Midas~\etal~\cite{Midas}. In this work, the additional baselines include:

3) \textbf{JITLine}~\cite{jitline}: A simple but effective method utilizing changed code and expert features to detect buggy commits. 

4) \textbf{JITFine}~\cite{jit-fine}: A DL-based approach extracting features of commits from changed code and commit message using CodeBERT as well as expert features.

Note that we did not utilize commit messages when adapting JITLine and JITFine for silent vulnerability fix identification in our experiments. For \tool, we set the number of GNN layers $L=2$ for a practical evaluation.

In this comparative study, we consider the impact of time on the approaches' performance.
Particularly, we divided the commits into those before and after the time point $t$. The commits before $t$ were used for training, while the commits after $t$ were used for evaluation. We selected a time point $t$ to achieve a random training/test split ratio of 80/20 based on time. Specifically, the commits from Aug 1998 to Mar 2017 are used for training, and the commits from Apr 2017 to Aug 2022 are for evaluation.

However, the vulnerability-fixing commits account for a very small proportion of the whole dataset, about 0.3\%. Consequently, the approaches work on a severely imbalanced classification dataset. This causes poor performance of classification approaches for the vulnerability detection task~\cite{resampling}. To mitigate this issue, Yang~\etal have recommended under-sampling for training classification models~\cite{resampling}. Thus, we applied under-sampling for all the considered approaches for a fair comparison.

For the testing dataset, the number of commits from Apr 2017 to Aug 2022 is huge, 2,462,900 commits, including only 2,828 vulnerability-fixing commits. We applied the approaches to the set containing all the fixing commits and the set of non-vulnerability-fixing commits manageable for all the considering approaches given our hardware resources. Specifically, we considered the testing dataset containing 2,828 vulnerability-fixing commits and about 80K+ non-fixing commits (Table~\ref{tab:dataset}).
Additionally, we applied the same procedure in the existing work by Zhou~\etal~\cite{devign} to project the relative comparison trend for the approaches with the real-world non-fix/fix ratio.

For \textbf{RQ2. Intrinsic Analysis}, we investigated the impact of the consideration of related unchanged parts and the GNN model on \tool's performance. We used different variants of \aAST and graph neural networks such as GAT~\cite{gat}, GCN~\cite{gcn}, GIN~\cite{gin}, and GraphSAGE~\cite{GraphSAGE} to study the impact of those factors on \tool's performance.

For \textbf{RQ3. Sensitivity Analysis}, we studied the impacts of the following factors on the performance of \tool: training size and change size in the number of changed LOC. To systematically vary these factors, we gradually added more training data and varied the range of the change size.

\subsection{Metrics}
Essentially, the task of vulnerability fix identification could be considered as a binary classification task. Thus, to evaluate the vulnerability fix identification approaches, we measure the classification \textit{accuracy}, \textit{precision}, and \textit{recall}, as well as \textit{F1}, which is a harmonic mean of precision and recall. 
Particularly, the classification accuracy (\textit{accuracy} for short) is the fraction of the (fixing and non-fixing) commits that are correctly classified among all the tested commits.
For detecting fixing commits, \textit{precision} is the fraction of correctly detected fixing commits among the detected fixing commits, while \textit{recall} is the fraction of correctly detected fixing commits among the fixing commits. Formally $precision = \frac{TP}{TP+FP}$ and $recall = \frac{TP}{TP+FN}$, where $TP$ is the number of true positives, $FP$ and $FN$ are the numbers of false positives and false negatives, respectively. \textit{F1} is calculated as $\textit{F1} = \frac{2 \times precision \times recall}{precision + recall}$.
Additionally, we also applied a cost-aware performance metric, \CEL (\textit{CE@L}), which is used in~\cite{Midas,VulFixMiner}. \textit{CE@L} counts the number of detected vulnerability-fixing commits, starting from commit with high to low predicted probabilities until the number of lines of code changes reaches $L$ lines of code (LOCs). In this work, we considered, $C@50K$,  $L = 50,000$, about 0.02\% of the total changed LOCs in our dataset, for simplicity.

\section{Experimental Results}
\label{sec:results}

\subsection{Performance Comparison (RQ1)}
Table~\ref{tab:comparision} shows the performance of \tool and the state-of-the-art vulnerability-fix identification approaches. As seen, \tool significantly outperforms the state-of-the-art vulnerability-fix identification approaches. 
Particularly, the \tool achieves a recall of 0.99. In other words, 99/100 vulnerability fixing commits are correctly identified by \tool, which is more than \textbf{22--70\%} better than the recall rates of the existing approaches. 
Additionally, \tool is still much more precise than the existing approaches with about \textbf{272--420\%} improvement in the precision rate. 
These show that \tool can not only find more vulnerability-fixing commits but also provide much more precise predictions.
Furthermore, the \textit{CE@50K} of \tool is \textbf{71\%}, which is \textbf{83-121\%} better than the corresponding figures of MiDas and VulFixMiner. This means that given the effort reviewing 50K LOC, the number of the fixing commits found by using \tool is much larger compared to those found by using MiDas and VulFixMiner.

\begin{table*}[]
\centering
\caption{Comparison Results}
\label{tab:comparision}
\begin{tabular}{@{}lrrrrrr@{}}
\toprule
 & \textit{Pre.} & \textit{Rec}. & \textit{F1} & \textit{Acc.} & \textit{AUC} & \textit{CE@50K} \\ \midrule
JITLine                         & 0.06 & 0.66 & 0.11 & 0.72 & 0.66      & 0.35  \\
JITFine                         & 0.07 & 0.81 & 0.13 & 0.65 & 0.80      & 0.39  \\ 
VulFixMiner                     & 0.05 & 0.05 & 0.05 & 0.62 & 0.65     & 0.32  \\
MiDas                           & 0.06 & 0.63 & 0.11 & 0.67 & 0.70      & 0.34  \\ 

\midrule

\tool            & \textbf{0.26} & \textbf{0.99} & \textbf{0.41} & \textbf{0.91} & \textbf{0.98} &  \textbf{0.71} \\ 
\bottomrule
\end{tabular}
\end{table*}

\begin{figure}
    \centering
    \includegraphics[width=1\columnwidth]{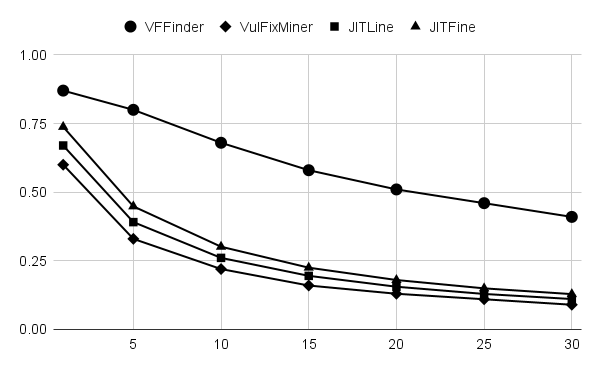}
    \caption{The performance of \tool and the existing approaches in different imbalance degrees (non-fix/fix rates)}
    \label{fig:label-rate}
\end{figure}

To project the approaches' performance on the real-world non-fix/fix rate, we investigated their performance on various datasets with different rates of fixing and non-fixing commits (Figure~\ref{fig:label-rate}). As seen, all the approaches perform worse when the proportion of non-fixing commits increases. \tool's F1 declines about 56\% when testing on the dataset with the non-fix/fix rates from 1:1 to 30:1. This phenomenon occurs for all the vulnerability-fix identification approaches. However, the performance of the existing approaches declines much faster than \tool's performance. Indeed, the other approaches' performance declines from  83--85\%. Furthermore, \tool performs much better than the existing approaches for all the considered rates.

Overall, \tool is more effective than the state-of-the-art approaches in identifying vulnerability fixes. This confirms our strategy explicitly representing the code structure changes and using graph-based models to extract features for vulnerability fix identification.

\subsection{Intrinsic Analysis (RQ2)}

To investigate the contribution of the related unchanged code in \aAST, we used two variants of \aAST: one considering both changed lines and unchanged lines (\aAST), the other considering only changed lines ($\widehat{\aAST}$). Table~\ref{tab:unchanged-code} shows the performance of \tool using the two \aAST variants: $\tool_\text{\aAST}$ and $\tool_{\widehat{\text{\aAST}}}$. 
For simplicity, in this experiment, we used the same GAT model for both representation variants and the dataset with the non-fix/fix rate of 1:1.

As seen, additionally considering the related unchanged lines along with changed lines in \aAST significantly improves the performance of \tool using \aAST with only changed lines. Particularly, $\tool_\text{\aAST}$ achieves an equivalent recall rate but a much better \textit{precision} rate, which is 21\% higher than that of $\tool_{\widehat{\text{\aAST}}}$. 
The related unchanged code provides valuable information and helps the model not only understand code changes more precisely but also discover more vulnerability-fix patterns. 
This experimentally confirms our observation 2 on the important role of related unchanged code for understanding code changes in vulnerability fix identification.

\begin{table*}[]
\centering
\caption{Impact of GNN Models}
\label{tab:unchanged-code}
\begin{tabular}{@{}llllll@{}}
\toprule
 & \textit{Pre.} & \textit{Rec}. & \textit{F1} & \textit{Acc.} & \textit{AUC} \\ \midrule

$\tool_\text{\aAST}$                & 0.8 & 0.97 & 0.88 & 0.87 & 0.85 \\
$\tool_{\widehat{\text{\aAST}}}$        & 0.66 & 0.97 & 0.78 & 0.73 & 0.74     \\
 \bottomrule
\end{tabular}
\end{table*}

To investigate the impact of different GNN models on the vulnerability-fix identification performance, we compare three variants of graph neural networks: \tool with GCN~\cite{gcn}, GAT~\cite{gat}, GIN~\cite{gin}, and GraphSAGE~\cite{GraphSAGE}. 
In this experiment, we use the dataset with the non-fix/fix rate of 1:1 for simplicity.
The results of those four variants are shown in Table~\ref{tab:gnn}. As expected, \tool obtains quite stable performance with the F1 of 0.88--0.91 and the accuracy of 0.87--0.90.
Moreover, while \tool obtains quite similar \textit{recall} with GCN, GAT, GIN, and GraphSAGE, it archives the highest \textit{precision}, yet lowest \textit{recall} with GAT.
%
%
%
Thus, lightweight relational graph neural networks such as GCN~\cite{gcn} should be applied to achieve cost-effective vulnerability-fix identification performance.

\begin{table*}[]
\centering
\caption{Impact of GNN Models}
\label{tab:gnn}
\begin{tabular}{@{}lrrrrr@{}}
\toprule
 & \textit{Pre.} & \textit{Rec}. & \textit{F1} & \textit{Acc.} & \textit{AUC} \\ \midrule
GCN~\cite{gcn}     & 0.84 & 0.98 & 0.90 & 0.90 & 0.90    \\
GAT~\cite{gat}     & 0.80 & 0.97 & 0.88 & 0.87 & 0.85   \\
GIN~\cite{gin}     & 0.81 & 0.99 & 0.89 & 0.88 &  0.91     \\
GraphSAGE~\cite{GraphSAGE}     & 0.85 & 0.97 & 0.91 & 0.90 & 0.91    \\ \bottomrule
\end{tabular}
\end{table*}

\subsection{Sensitivity Analysis (RQ3)}

\begin{figure}
    \centering
    \includegraphics[width=1\columnwidth]{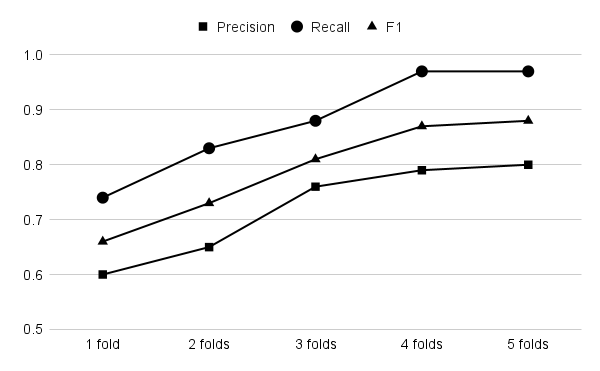}
    \caption{Impact of training data size on \tool's performance}
    \label{fig:impact_training_size}
\end{figure}

To measure the impact of training data size on \tool's performance. In this experiment, the training set is randomly separated into five folds. We gradually increased the training data size by adding one fold at a time until all five folds were added for training. Note that, in this experiment, we used GAT and the dataset with the non-fix/fix rate of 1:1.
As shown in Figure~\ref{fig:impact_training_size}, \tool's performance is improved when expanding the training dataset. The precision, recall, and F1 increase by more than 30\% when the training data expands from one fold to five folds. The reason is that with larger training datasets, \tool has observed more and performs better. However, training with a larger dataset costs more time. The training time of \tool with five folds is about 4.53 more than that with a fold.

\begin{figure}
    \centering
    \includegraphics[width=1\columnwidth]{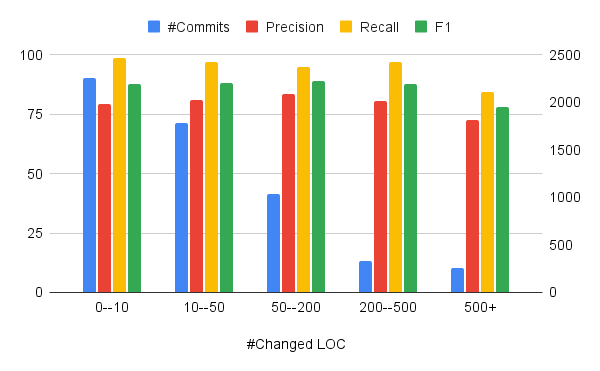}
    \caption{Impact of change size (left axis: \textit{Precision} and \textit{Recall}; right axis: No. of commits)}
    \label{fig:impact_change_size}
\end{figure}

Additionally, we investigate the sensitivity of \tool's performance on the input size in the number of changed (i.e., added and deleted) lines of code (LOCs) (Fig.~\ref{fig:impact_change_size}). As seen, there are much fewer commits with a larger number of changed LOCs. 
The \textit{precision} of \tool is quite stable when handling commits in different change sizes. Particularly, \tool's F1 slightly varies between 0.88 and 0.89 when increasing the change size from 1 to 500 changed LOC. 
For the fixing commits changing a large part of code, \tool can still effectively identify with the recall of 0.73 and the precision of 0.84. Additionally, \tool's F1 for the vulnerability fixing commits having more than 500 changed LOC is only about 10\% lower than \tool's performance for other vulnerability fixing commits.

\subsection{Time Complexity (RQ3)}
In this work, all our experiments were run on a server running Ubuntu 18.04 with an NVIDIA Tesla P100 GPU. 
In \tool, training the model took about 4--6 hours for 50 epochs. Additionally, \tool spent 1--2 seconds to classify whether a commit is a fixing commit or not. 

\subsection{Threats to Validity}
The main threats to the validity of our work consist of internal, construct, and external threats.

\textbf{Threats to internal validity} encompass the potential impact of the adopted method for building Abstract Syntax Trees (ASTs). To mitigate this challenge, we employ the extensively recognized code analyzer Joern~\cite{joern}. Another threat lies in the correctness of the implementation of our approach. To reduce such a threat, we carefully reviewed our code and made it public~\cite{VFFinder} so that other researchers can double-check and reproduce our experiments.

\textbf{Threats to construct validity} relate to the suitability of our evaluation procedure. 
We used \textit{precision}, \textit{recall}, \textit{F1}, \textit{AUC}, \textit{accuracy}, and \CEL. 
They are the widely-used evaluation measures for vulnerability fix identification and just-in-time defect detection~\cite{jit-fine,jitline, VulFixMiner, Midas}. 
Besides, a threat may come from the adaptation of the baselines. To mitigate this threat, we directly obtain the original source code from their GitHub repositories or replicate exactly their description in the paper~\cite{VulFixMiner,Midas}. Also, we use the same hyperparameters as in the original papers~\cite{jit-fine,jitline,deepjit,Midas}. 

\textbf{Threats to external validity} mainly lie in the selection of graph neural network models employed in our experiments. 
To mitigate this threat, we have chosen widely recognized models with established track records in NLP and SE domains~\cite{gat, gcn,GraphSAGE,gin}. 
%
Moreover, our experiments are conducted on only the code changes of C/C++ projects. Thus, the outcomes may not be universally applicable to different programming languages. To overcome this limitation, our future research agenda involves performing additional experiments to validate the findings across various programming languages.
\section{Related Work}
\label{sec:related}









\tool relates to the vulnerability fix identification work. VulFixMiner~\cite{VulFixMiner} and CoLeFunDa~\cite{CoLeFunDa} utilize CodeBERT to automatically represent code changes and extract features for identifying vulnerability-fixing commits.
Midas~\cite{Midas} constructs different neural networks for each level of code change granularity, corresponding to commit-level, file-level, hunk-level, and line-level, following their natural organization. It then utilizes an ensemble model that combines all base models to generate the final prediction.

\tool also relates to the work on just-in-time vulnerability detection.
DeepJIT~\cite{deepjit} automatically extracts features from commit messages and changed code and uses them to identify defects. 
Pornprasit~\etal propose JITLine, a simple but effective just-in-time defect prediction approach. JITLine utilizes the expert features and token features using bag-of-words from commit messages and changed code to build a defect prediction model with a random forest classifier. 
LAPredict~\cite{lapredict} is a defect prediction model by leveraging the information of ``lines of code added'' expert feature with the traditional logistic regression classifier. 
Recently, Ni~\etal introduced JITFine~\cite{jit-fine}, combining the expert features and the semantic features which are extracted by CodeBERT~\cite{codebert} from changed code and commit messages. 
Different from all prior studies in vulnerability fix identification and just-in-time bug detection, our work presents \tool which explicitly represents code changes in code structure and applies a graph-based model to extract the features distinguishing fixing commits from non-fixing ones. 


Several studies have been proposed for specific SE tasks, including code suggestion/completion~\cite{icse20, autosc,arist}, program synthesis~\cite{gvero2015synthesizing}, pull request description generation~\cite{hu2018deep,liu2019automatic}, code summarization~\cite{iyer2016summarizing,mastropaolo2021studying,wan2018improving}, code clones~\cite{li2017cclearner}, fuzz testing\cite{godefroid2017learn}, code-text translation~\cite{ase22}, bug/vulnerability detection~\cite{oppsla19, codejit, vultype}, and program repair~\cite{jiang2021cure,ding2020patching}.
\section{Conclusion}
\label{sec:conclusion}

In conclusion, this paper has addressed the critical challenge of identifying silent vulnerability fixes in software projects heavily reliant on third-party libraries. The existing gap between fixes and public disclosures, coupled with the prevalence of undisclosed vulnerability fixes in open-source projects, has hindered effective vulnerability management.
In this paper, we have introduced \tool, a novel graph-based approach designed for the automated identification of vulnerability-fixing commits. To precisely capture the meaning of code changes, the changed code is represented in connection with the related unchanged code. To precisely capture the meaning of code changes, the changed code is represented in connection with the related unchanged code. In \tool, the structure of the changed code and related unchanged code are captured and the structural changes are represented in annotated Abstract Syntax Trees. By leveraging annotated ASTs to capture structural changes, \tool enables the extraction of essential structural features. These features are then utilized by graph-based neural network models to differentiate vulnerability-fixing commits from non-fixing ones.
Our experimental results show that \tool improves the state-of-the-art methods by 272--420\% in Precision, 22--70\% in Recall, and 3.2X--8.2X in F1. Especially, \tool speeds up the silent fix identification process by up to 121\% with the same effort reviewing 50K LOC compared to the existing approaches.
These findings highlight the superiority of \tool in accurately identifying vulnerability fixes and its ability to expedite the review process. The performance of \tool contributes to enhancing software security by empowering developers and security auditors with a reliable and efficient tool for identifying and addressing vulnerabilities in a timely manner.

\bibliographystyle{IEEEtran}

\bibliography{main}

\end{document}